\documentclass[preprint]{aastex}

\usepackage{natbib}
\usepackage{gensymb}
\shorttitle{Evolution of Near-surface Flows}
\shortauthors{Bogart, Baldner, \& Basu}

\begin{document}
\title{Evolution of Near-surface Flows Inferred from High-resolution
  Ring-diagram Analysis}
\author{Richard S.~Bogart, Charles S.~Baldner}
\affil{Stanford University, Stanford, CA 94305-4085, USA}
\and
\author{Sarbani Basu}
\affil{Yale University, New Haven, CT 06520, USA}

\begin{abstract}
Ring-diagram analysis of acoustic waves observed at the photosphere can
provide a relatively robust determination of the sub-surface flows at
a particular time under a particular region. The depth of penetration of the
waves is related to the size of the region, hence the depth extent of the
measured flows is inversely proportional to the spatial resolution. Most
ring-diagram analysis has focused on regions of extent $\sim$15\degree~(180 Mm) or more in
order to provide reasonable mode sets for inversions. HMI data analysis
also provides a set of ring fit parameters on a scale three
times smaller. These provide flow estimates for the outer 1\% (7 Mm) of
the Sun only, with very limited depth resolution, but with spatial resolution
adequate to map structures potentially associated with the belts and regions
of magnetic activity. There are a number of systematic effects affecting the
determination of flows from local helioseismic analysis of regions over
different parts of the observable disk, not all well understood. In this
study we characterize those systematic effects with higher spatial resolution,
so that they may more effectively be accounted for in mapping temporal and
spatial evolution of the flows. Leaving open the question of the mean
structure of the global meridional circulation and the differential rotation,
we describe the near-surface flow anomalies in time and latitude corresponding
to the torsional oscillation pattern in differential rotation and analogous
patterns in the meridional cell structure as observed by SDO/HMI.
\end{abstract}

\section{Introduction}
The mean properties of solar rotation, zonal flows and meridional circulation
close to the surface have been well studied. The Sun rotates substantially
faster at the equator than at the poles. Over and above the mean differential
rotation profile  there are bands of mean zonal flows corresponding to slight
prograde and retrograde variations in the local rotation rate at
different latitudes that change with the solar cycle. There is also
a regular weak flow of material from the solar equator to the poles.
The most reliable measurements of these dynamical features are
for the low- and mid-latitude regions. Limitations of observation and analysis
techniques have resulted in rather uncertain results at high solar latitudes.
There are also uncertainties in the details of the longitudinal variation of
solar dynamics. Observations made by the Helioseismic and Magnetic
Imager (HMI) on board the Solar Dynamics Observatory (SDO) allow us to examine
both of these issues. In this paper we use these data to study the evolution
of spatially resolved features in the zonal and meridional flows at
high solar latitudes.

The presence of zonal flows was first deduced from surface Doppler measurements.
A long-period oscillation, dubbed the ``torsional oscillation,'' was identified
by \citet{HowardLaBonte1980}. It consists of alternating latitude bands of
slightly faster and slower rotation
migrating equatorward as the solar activity cycle progresses. These were
confirmed by \citet{Ulrich1998, Ulrich2001}. Poleward meridional circulation
was detected at the surface using direct Doppler measurements
\citep{LaBonteHoward1982,Hathawayetal1996}.

Both zonal and meridional flows seen at the surface have been confirmed
by various helioseismic studies. The longitudinally averaged, north-south
symmetric part of the zonal flows has been seen by inverting global modes of
solar oscillations (e.g. \citep[e.g.][]{KosovichevSchou1997, antiabasu2000,
howeetal2000, antiabasu2013, howeetal2013}.
The lack of sensitivity of the global data to features at very
high latitudes means that these investigations could not probe the polar
regions properly. Additionally, global helioseismic inversions do not give
any longitudinal resolution, nor can they readily separate the contributions
from the northern and southern hemispheres. Local helioseismic techniques,
however, can address these problems.

Among the more widely used local helioseismic techniques,
ring-diagram analysis \citep{Patronetal1997,Hill1988} uses
three-dimensional power spectra of small regions of the Sun to
measure the mean horizontal motions over those regions, allowing us
to study localized solar dynamic phenomena.  The analysis technique has
allowed us
to study hemispherical differences in the near-surface rotation rate
\citep[e.g.][]{Basuetal999,Haberetal2000}, as well as meridional flows
\citep[e.g.][]{Basuetal999, Haberetal2002, BasuAntia2010}.

Determining the exact nature of the meridional flows at high latitudes has
proved to be a challenge. At low and
middle latitudes, the flow of material is poleward in both hemispheres. The
situation at high latitudes is less clear --- even the sense of the flow is
difficult to determine. Equatorward cells at very high latitudes have been
reported, but it has been shown that these may be due, at least in part, to
systematic errors in the measurements at large distances from disk center
\citep{Zaatrietal2006,Zhao2012}.

The  high-resolution, high-cadence, nearly continuous data available for the
current solar cycle from the HMI instrument make it possible to study
spatially resolved flows up to very high latitudes. Surface Doppler data were
used by \citet{Hathaway13}
to study the motion of supergranules; they found long-lived anomalies of
slower-than-average (retrograde) zonal flows, which they interpreted as
the signatures of giant cell convection. Using the HMI Doppler data for
local helioseismic analysis of solar near-surface dynamics, we can
see similar long-lived features close to the poles. We use
data collected over a four-year period from 2010 to 2014 that covers the rising
phase of cycle 24. We use the ring-diagram technique, and unlike prior
investigations such as those of \citet{Kommetal2013} and 2015, we use tiles as small as $5^\circ\times 5^\circ$ instead of the
usual $15^\circ\times 15^\circ$ to achieve better spatial resolution.

In Section 2 we discuss the data analysis. Detailed results are presented in
Section 3, and we conclude with a summary of the results in Section 4.

\section{Data Analysis}
The Helioseismic and Magnetic Imager (HMI) has been producing a nearly
uninterrupted series of full-disk solar photospheric Dopplergrams suitable
for high-resolution local helioseismic analysis at a cadence of one every
45 sec for over 4.5 years. One of its fundamental mission science products
is a collection of ``ring-diagram'' analysis products --- power spectra,
ring (ridge) fit parameters, and depth
inversions of certain of the horizontal flow parameters
\citep{Bogart11a,Bogart11b}.
The ring analyses are performed on three different scales, for
sets of overlapping regions of approximate diameters 5\degree, 15\degree,
and 30\degree\ heliographic,
or about 60, 180, and 360 Mm respectively. Most
published work has focused on the data for the 15\degree~regions, in part
because this scale is also accessible to the lower-resolution data from
the Michelson Doppler Imager (MDI) on the earlier SoHO mission and the
ongoing GONG project \citep[see][and references therein]{ringsrev13}. 
Furthermore, most of the studies of the evolution of synoptic flow structures
have been limited to data from regions on the central meridian. As we
show here, the quality of the mode fits is nearly uniform over most of
the solar disk, suggesting that mode fits from off the meridian may
contribute substantially to the quality of the measurements, particularly
over short time samples.

The data for the 5\degree~regions do not yield mode sets rich enough to support
well-resolved depth inversions, as they are normally sensitive to only
the $f$-mode and two or three orders of the $p$-modes, penetrating to depths
of about 1\% (7 Mm) of the solar radius (see Figure 1).
They do however provide spatial
resolution adequate to map structures potentially associated with the
zonal belts and regions of magnetic activity. They also provide fine
enough spatial resolution to address questions about systematic disk-position
effects which may bias measurements of the structure of the large-scale
mean flows \citep[e.g.][]{Zhao2012}. These systematic effects are of
particular concern for measurement of the meridional circulation at
high latitudes, an important parameter for models of solar cycle
evolution. As we show here, these effects do not have a simple radial
nor east-west dependence. Nevertheless, they appear to be quite stable
over time. This suggests that they have little or no effect on the
short-term evolution of the flow patterns, which are the focus of this
study.

In the HMI ring-diagram analysis pipeline, sequences of Dopplergrams are
mapped into sets of tiles of equal size and shape, with each tile fixed at
particular Carrington coordinates ({\sl i.e.} tracked at the Carrington rotation
rate). Three different tile sizes are mapped, the tiles being tracked for
approximately the time during which they would rotate through their width.
The smallest tiles are mapped at the equivalent full resolution of HMI at
disk center, 0\degree.04 heliographic per pixel, with widths of 5\degree.12,
and tracked
for 576 min (768 HMI 45-sec Dopplergams). The tile centers are spaced at
intervals of 2\degree.5 in latitude and multiples of 2\degree.5 in longitude,
so that
they are roughly equally spaced on the surface, and extend to about
80\degree\ from
disk center. There are either 2727 or 2748 such tiles at each analysis time,
depending on $B_0$, the heliographic latitude of Earth. The analyses
are repeated at intervals of 5\degree\ in synodic rotation, 72 per Carrington
rotation. Apart from occasional interruptions in the data stream caused
by eclipses, instrument calibrations, spacecraft maneuvers, and accidents,
there are between 196,344 and 197,856 ring-diagram spectra analyzed per
rotation.

Two procedures are used to fit the power spectrum to obtain the mode
parameters. A fast algorithm designed to estimate just the frequency
displacements necessary for determinations of the mean horizontal flow
is applied to each region; its fit parameters are inverted to obtain the
published flows for the larger regions. For the 5\degree\ regions, however, the
number of modes fit is too small for useful inversions. Better fits are
obtained using the model described by \citet{BAB2004}:
\begin{eqnarray}
&&\hspace{-40 pt}P(k_x,k_y,\nu)= {e^{B_1}\over k^3} +{e^{B_2}\over k^4
}+\nonumber\\
&&\hspace{-30 pt}{\exp(A_0+(k-k_0)A_1+A_2({k_x\over k})^2+
A_3{k_xk_y\over k^2})S_x\over x^2+1}
\end{eqnarray}
where
\begin{eqnarray}
x&=&{\nu-ck^p-U_xk_x-U_yk_y\over w_0+w_1(k-k_0)},\\
S_x&=&S^2+(1+Sx)^2,
\end{eqnarray}
The 13 parameters
$A_0$, $A_1$, $A_2$, $A_3$, $c$, $p$, $U_x$, $U_y$, $w_0$, $w_1$, $S$, $B_1$,
and $B_2$
are determined by independently fitting each ridge in the spectrum.
The parameters of most interest are $U_x$ and $U_y$, representing the
mean advection in the zonal and meridional directions in the local
plane geometry at the center of the region.

These fits are too time consuming to perform over all of the larger regions,
but they can be and are applied to all of the 5\degree\ regions. Because they
provide
robust estimates of all of the mode parameters over almost the entire disk
with minimal cross-talk and fairly uniform precision, they are the ones
we have chosen to use for this study. The number of ridges that can be
fit is still of course limited by the small size of the regions; typically
only the $f$-mode and the first 3 $p$-mode ridges yield substantial numbers
of modes (Figure 1).
However, the fitting is of nearly uniform quality over the disk,
the number of modes fit among the lowest 4 orders being approximately
constant from disk center to about $\mu$ = 0.3 (72\degree.5), as shown in
Figure 2.
For estimation of
azimuthal averages of the flows, use of the fits from off of the central
meridian increases the available data by a factor of 42.

\begin{figure}
  \scalebox{0.65}
  {\includegraphics[angle=-90]{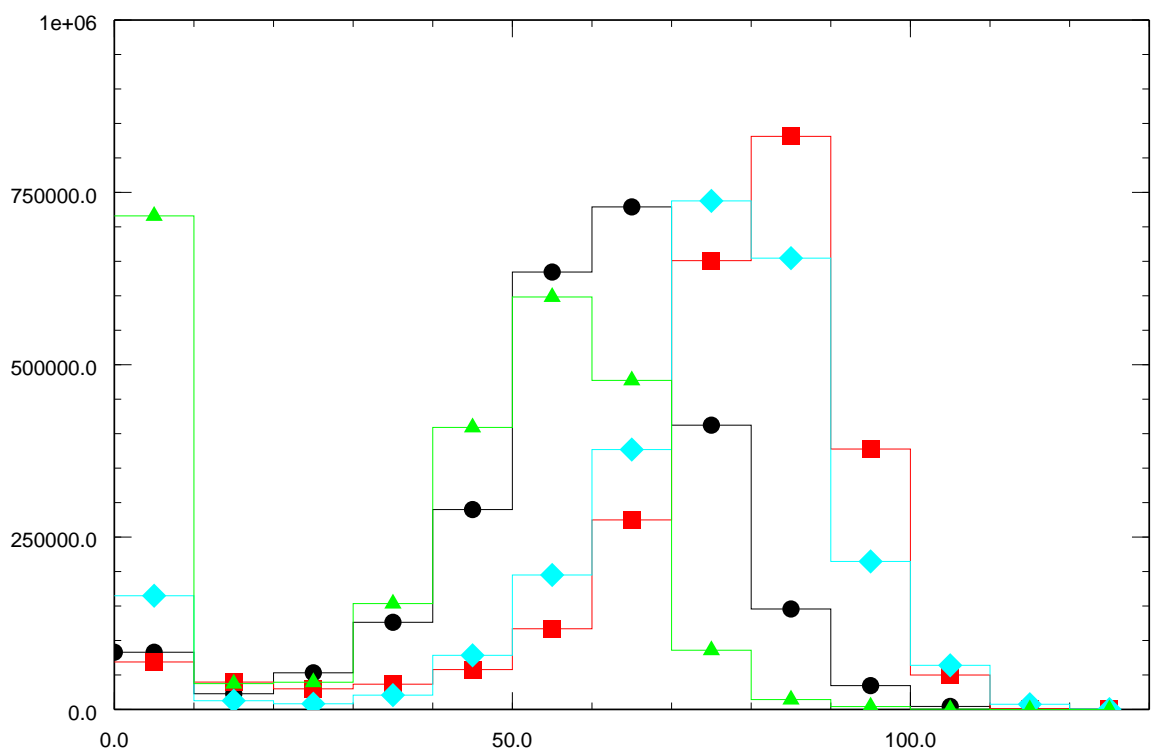}}
\caption{Histogram of the total number of regions with the given number of
mode fits, by radial order $n$, during a period of about one year (CR 2143--2155,
2013 October 25 -- 2014 October 14). Symbol codes: $n=0$ black circles,
$n=1$ red squares, $n=2$ cyan lozenges, $n=3$ green triangles.}
\label{fig:fitshist}
\end{figure}

\begin{figure}
  \scalebox{0.65}
  {\includegraphics[angle=-90]{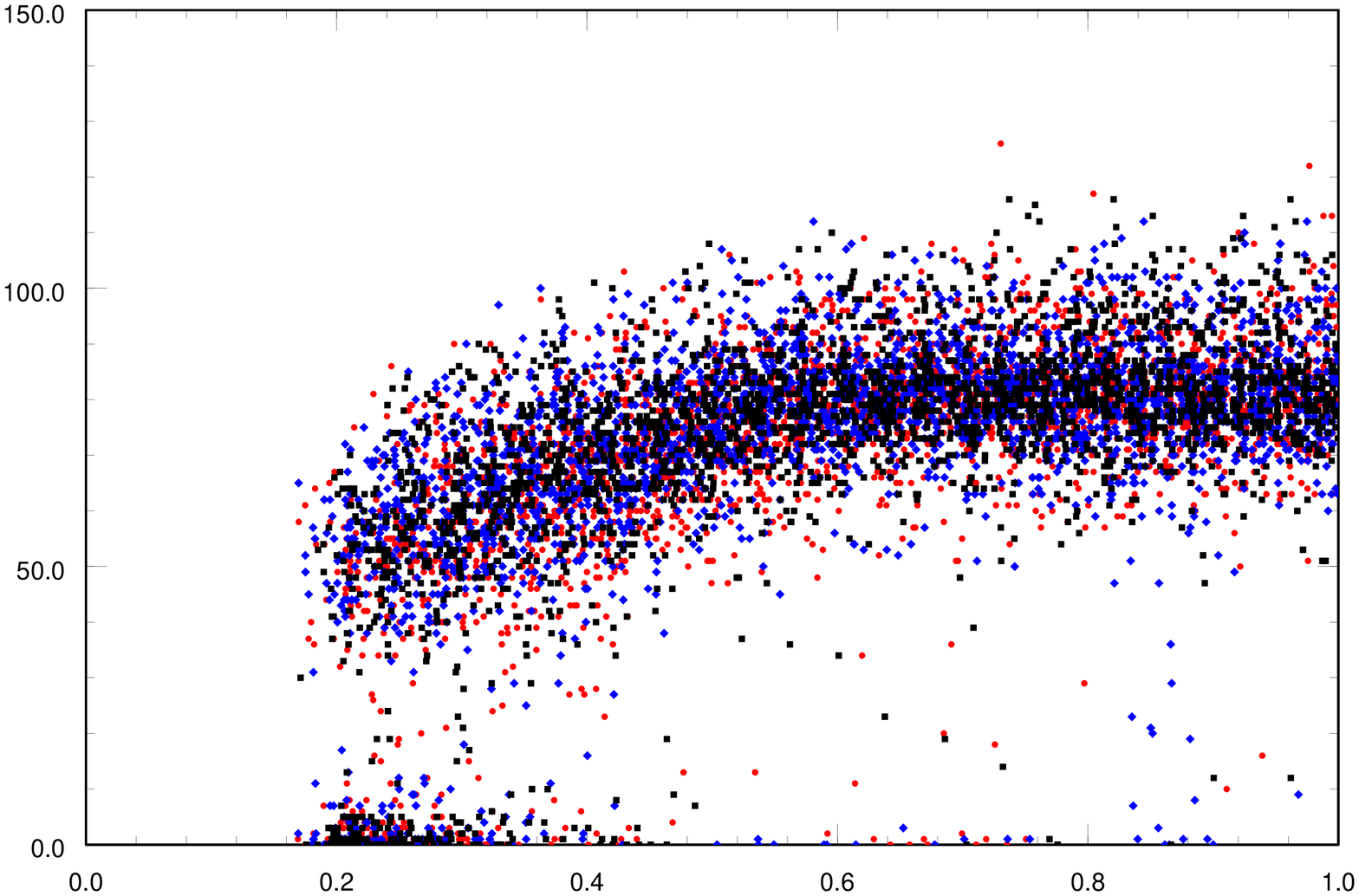}}
\caption{Scatter plot of the number of fits of order $n=2$ modes over the
entire disc as a function of the mean central angle of the center of the
region $\mu = \cos(\theta)$ as it is tracked during three sample time
intervals: 2121:340
(2012 March 06, when $B_0 = -7\degree.25$, red circles), 2154:115 (2014 September09,
when $B_0 = +7\degree.25$, blue lozenges), and 2144:160 (2013 December 06, when
$B_0 = 0\degree$, black squares).}
\label{fig:n2scatplots}
\end{figure}

Note that there is a systematic change in the $U_i$ parameter values between
the fits for Carrington Rotations 2125 and 2126 ({\sl i.e.} beginning
2012 July 18).
Analysis of data from the transit of Venus on 2012 June 5--6 
revealed that the original estimate for the HMI position angle with
respect to the Carrington rotation axis was slightly in error.
An adjustment of 0\degree.070206 to the position angle in processing of the
Doppler and other data was made beginning 2012 August 31 (actually at
23:48 on August 30) and the records for all data prior to that time
subsequently adjusted to reflect the correction. Data through the end
of CR 2125 had however already been processed using the old values of
the position angle for the mapping and tracking. There is thus a spurious
meridional component of the solar rotation, plainly visible in comparing
results from before and after the change
(see {Figure 7}; also {Figure 12}),
biasing the measured meridional flows by about 2.5 m-s$^{-1}$ in the
earlier data.

In order to examine temporal variations in the ``mean'' flow patterns, we
have divided the data into eight overlapping sets as close to one year in
length as possible (Table 1), in order to remove systematic biases due to
the annual variation of the observer latitude of $\pm7\degree.25$. Because
of the aforementioned adjustment in the assumed image position angles,
averages for Years 2.5 and 3 will include data with both values and
consequently mixed leakages of rotation into the mean meridional flow.

\begin{tabular}{|c c c c c|}
  \hline
  {\bf Year} & Start CR:CL & Start Time (UT) & End CR:CL & End Time (UT) \\
  \hline
  1 & 2096:250 & 2010 May 01 02:12 & 2109:115 & 2011 Apr 30 21:34 \\
  1.5 & 2102:010 & 2010 Oct 29 15:58 & 2116:235 & 2011 Oct 29 11:18 \\
  2 & 2109:110 & 2011 May 01 06:39 & 2123:335 & 2012 Apr 30 02:01 \\
  2.5 & 2116:230 & 2011 Oct 29 20:24 & 2129:095 & 2012 Oct 28 15:45 \\
  3 & 2123:330 & 2012 Apr 30 11:06 & 2136:190 & 2013 Apr 30 15:32 \\
  3.5 & 2129:090 & 2012 Oct 29 00:51 & 2143:310 & 2013 Oct 29 05:17 \\
  4 & 2136:185 & 2013 May 01 00:37 & 2149:050 & 2014 Apr 30 19:59 \\
  4.5 & 2143:305 & 2013 Oct 29 14:23 & 2156:170 & 2014 Oct 29 09:43 \\
  \hline
\end{tabular}

\section{Results}

When we examine the mean values of the $U_i$ parameters at each latitude
and (Stonyhurst) longitude averaged over exactly four years we see that (a) there are systematic and significant east-west
variations in both the zonal and meridional flows at a given latitude; and (b)
there is evidence for a consistent meridional flow reversal poleward of about
$\pm60\degree$ ({Figures 3 \& 4}). It is somewhat difficult to compare
the systematic disc variations in the zonal velocity that we see with those
from the similar lower spatial resolution analyses reported by
\citet{Kommetal2015} because they compare the systematic anomalies
with respect to a differential rotation model; but in any case the main
variations that we see occur at center-to-limb distances beyond the range
of their data. The azimuthal variations in the mean meridional flow that
we see exhibit a similar structure in the northern hemisphere to those
that they see at the times when the solar north pole is tipped towards
the Earth; but the systematic disc variations that we see are of much
lower amplitude. Again, the latitudes where our data show consistent evidence
for a high-latitude reversal are at and beyond the limit of theirs.

There are also small but significant year-to-year
variations in the meridional flow profile around the latitudes of activity.
The large zonally symmetric component to the zonal flow parameter merely
reflects the fact that the analysis regions are tracked at a uniform rate
to account for the rotation of the coordinate system, and consequently
exhibits the near-surface differential rotation in the outer 1\% of the
Sun, the average sampled depth.
East-west variations must certainly be spurious artifacts of either the
observational or analysis procedures, possibly related to center-to-limb
variations or the symmetry breaking associated with tracking the analysis
regions to compensate for the rotation of the coordinate system. The former
at least calls into question the reality of the observed high-latitude
counter cell in the meridional flow. The arguments advanced by Zhao {\sl et al.}
(2012) for rejecting a comparable result in time-distance analysis do not
however apply in this case, as significant limb effects in the mean zonal flow
parameter do not show up until about longitude 75\degree\ ({Figure 5}).
We also see small but significant hemispheric asymmetries in the mean zonal
flow (rotational) profile at the level of a few m-s$^{-1}$ at all latitudes between
about 10\degree and 75\degree, with faster motion in the southern hemisphere,
particularly around latitude $-50\degree$ ({Figure 6}).

\begin{figure}
  \scalebox{0.75}
  {
    \includegraphics{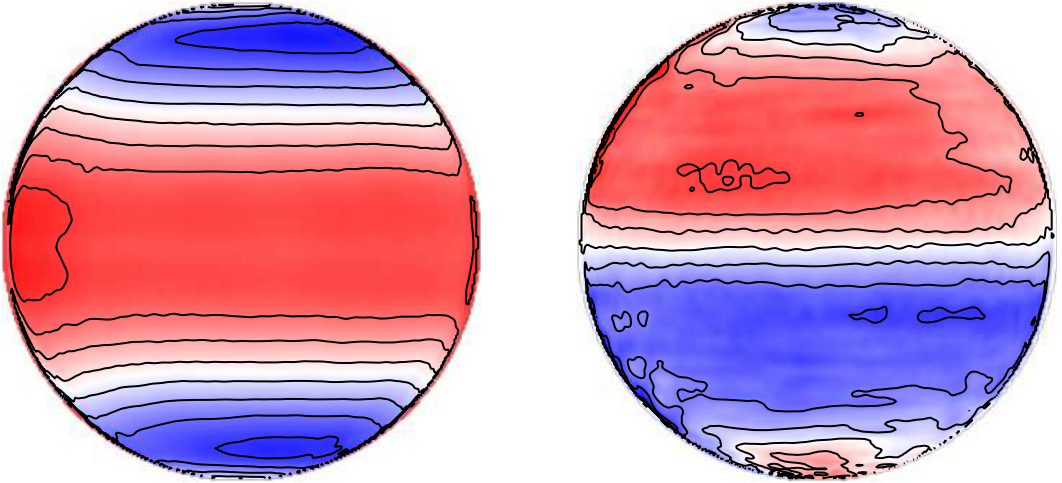}
  }
  \caption{
  Mean values of $U_x$ (left) and $U_y$ (right) over the 4-year interval
  2096:250--2149:050 (2010 May 1 to  2014 May 1) integrated over all
  modes in the range $n=\hbox{1--3}$ with $R_t$ between 0.9875 and 0.9975, as
  functions of heliographic latitude and Stonyhurst longitude. The color scales
  range from -200 -- +25 m-s$^{-1}$ with 25 m-s$^{-1}$ contours for $U_x$, and
  between $\pm20$ m-s$^{-1}$ with 5 m-s$^{-1}$ contours for $U_y$.}
\label{fig:mean4yrdisk}
\end{figure}

\begin{figure}
  \scalebox{0.325}
  {
   \includegraphics[angle=-90]{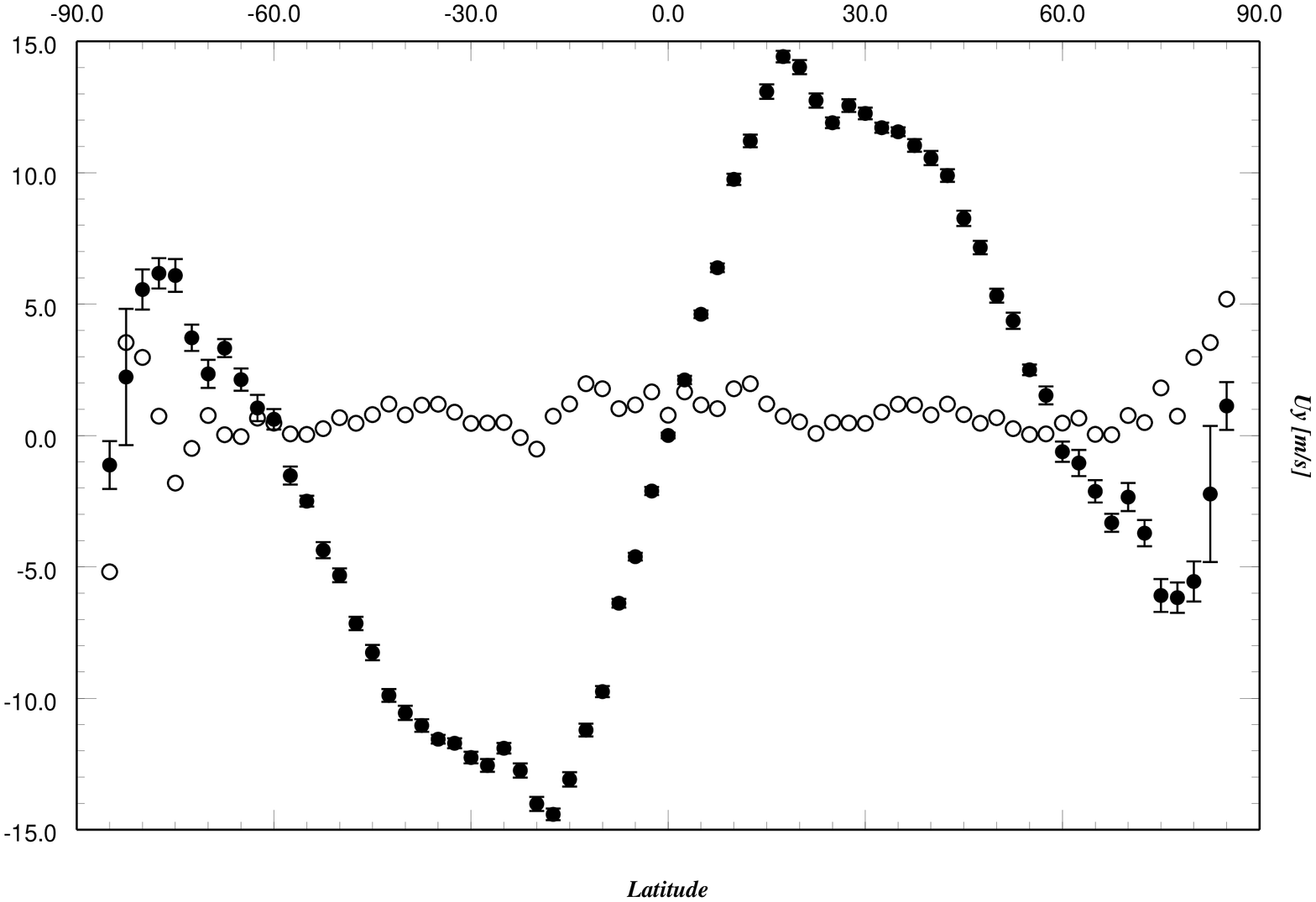}
   \includegraphics[angle=-90]{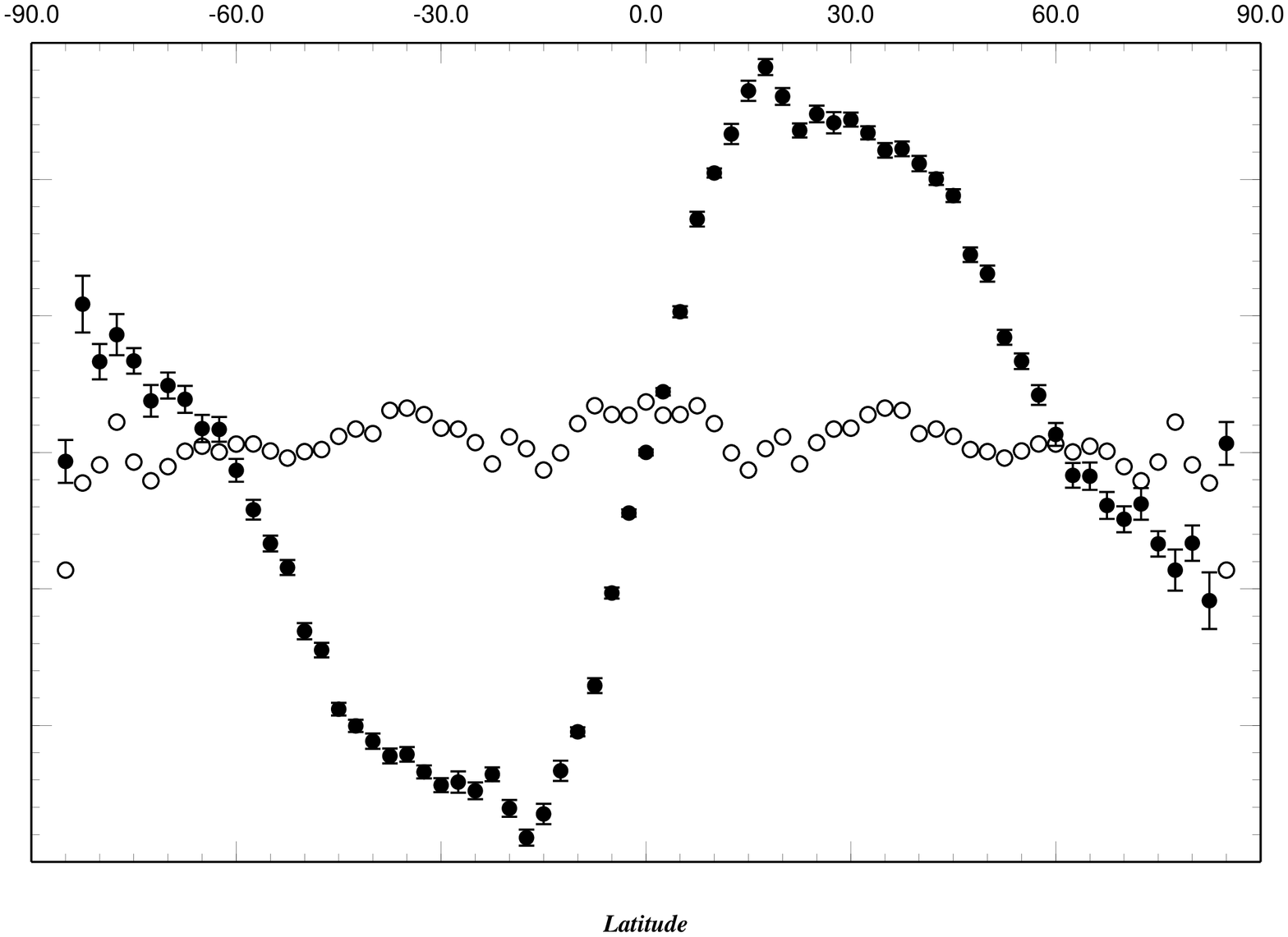}
  }
  \scalebox{0.325}
  {
   \includegraphics[angle=-90]{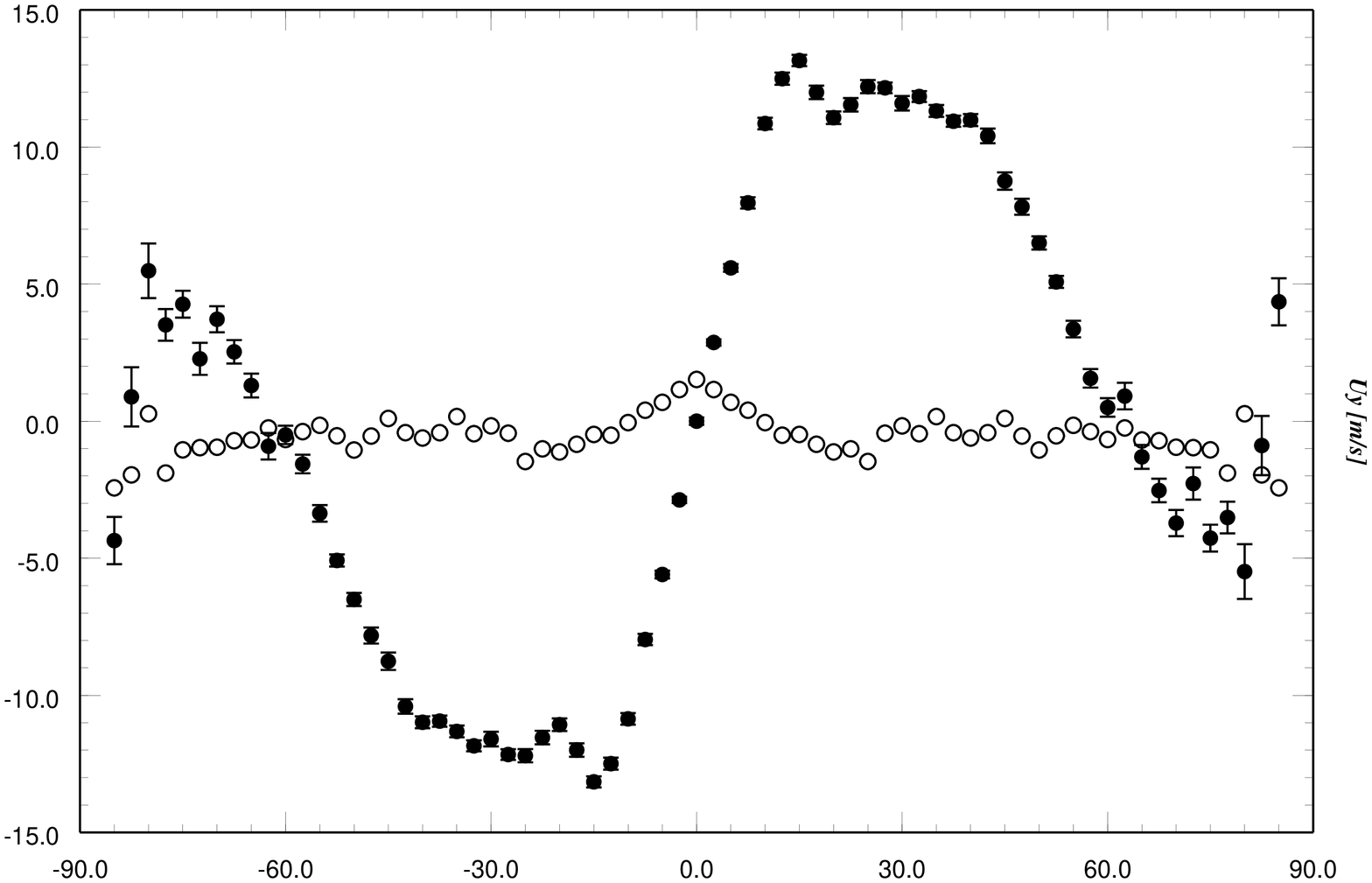}
   \includegraphics[angle=-90]{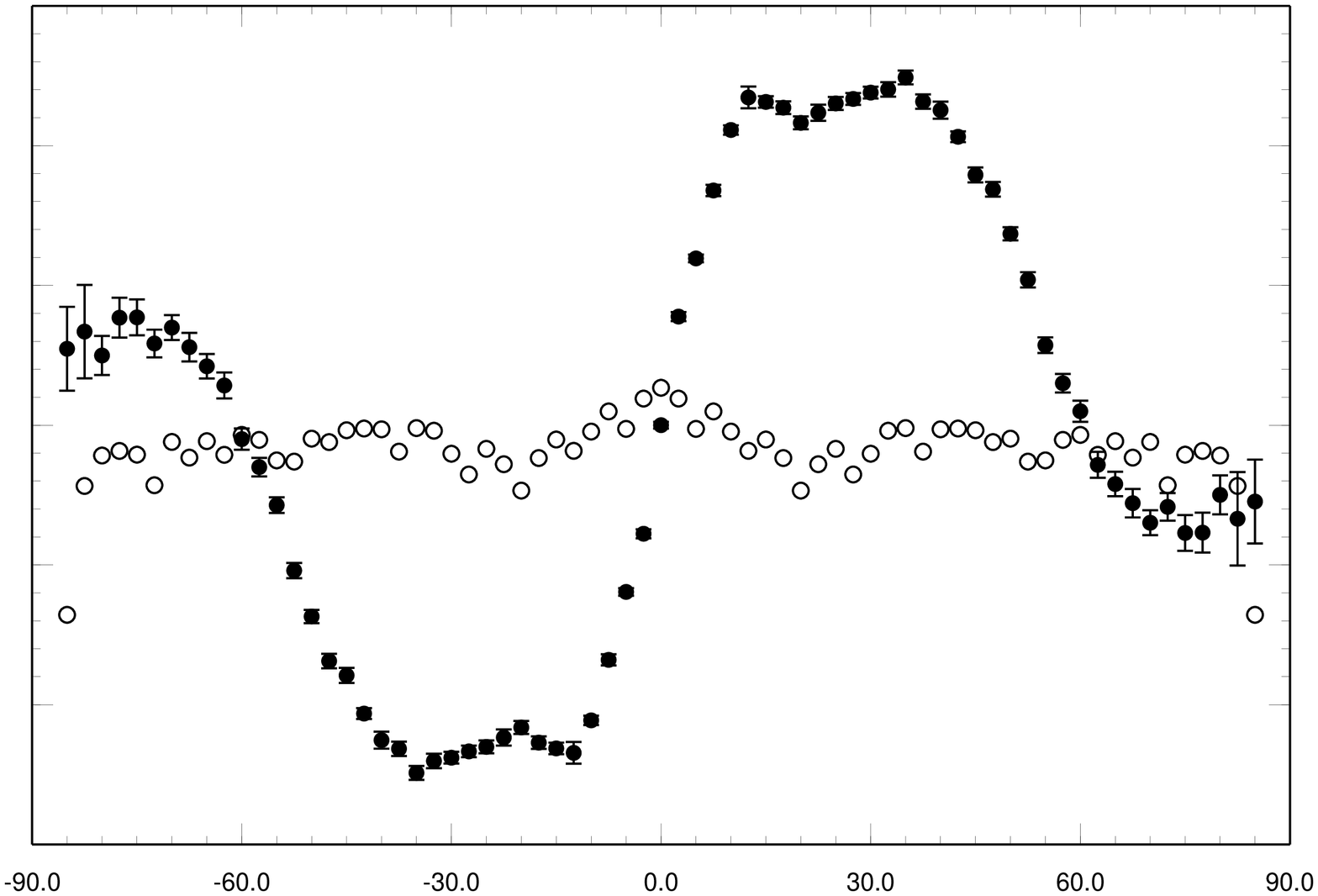}
  }
  \caption{
  Zonally symmetric (open symbols) and anti-symmetric (filled symbols)
  components of the meridional flow parameter $U_y$ averaged over all
  longitudes for the same mode sets as in Fig. 3 for Years 1 (upper left),
  2 (upper right), 3 (lower left), and 4 (lower right).
  Standard error bars are shown for the anti-symmetric component only; those
  for the symmetric components at the same latitude are of course the same.}
\label{fig:uycomp}
\end{figure}

\begin{figure}
  \scalebox{0.65}
  {\includegraphics[angle=-90]{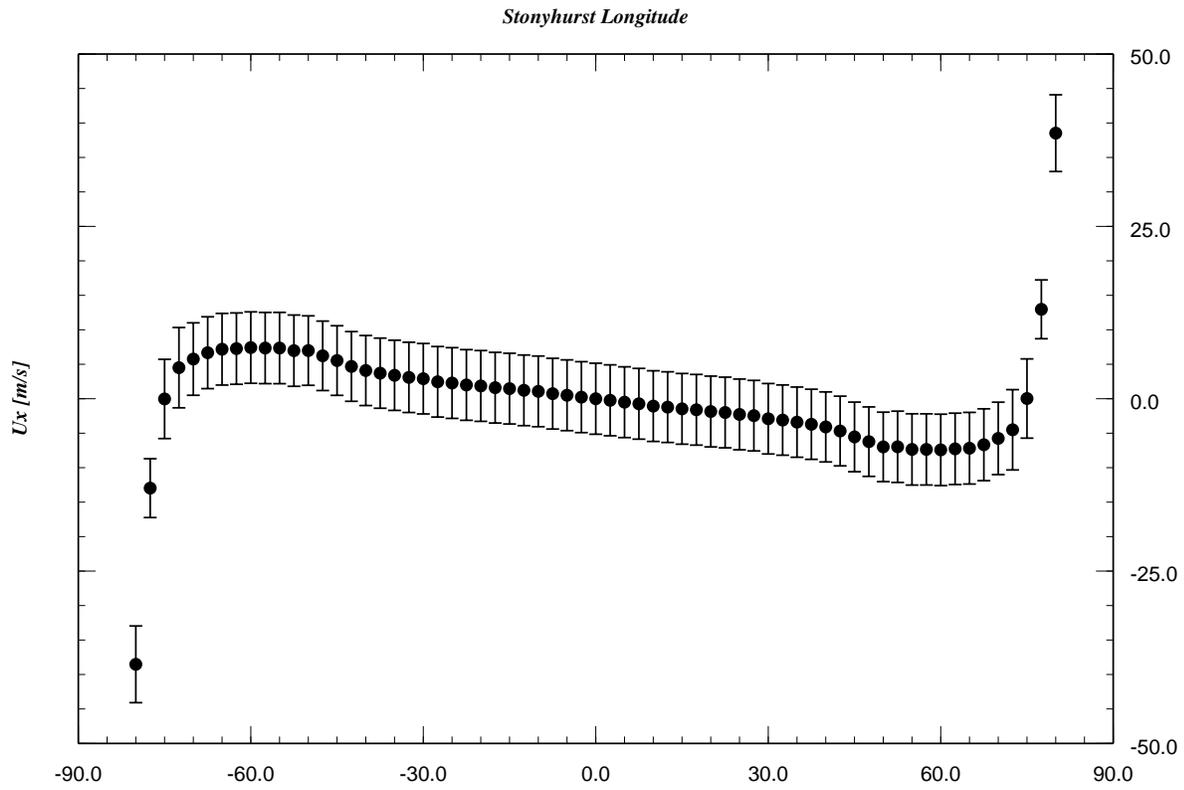}}
  \caption{
  Longitudinally antisymmetric component of the 4-year averaged values of the
  zonal flow parameter $U_x$ averaged over all latitudes for the same mode
  sets as in Fig. 3, shown as a funtion of Stonyhurst longitude.}
\label{fig:uxmanti}
\end{figure}

\begin{figure}
  \scalebox{0.65}
  {\includegraphics[angle=-90]{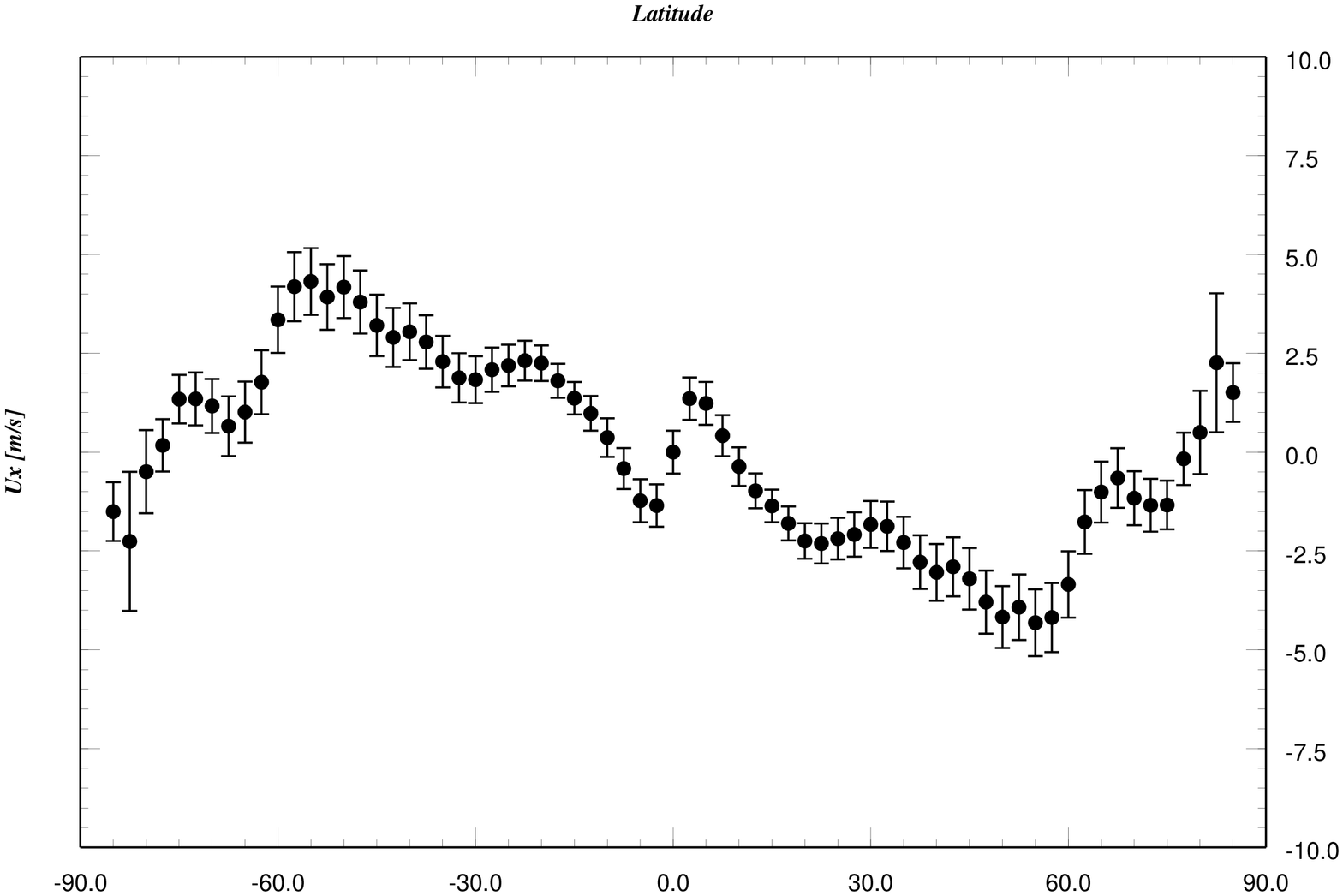}}
  \caption{
  Zonally antisymmetric component of the 4-year averaged values of the zonal flow
  parameter $U_x$ averaged over all longitudes within $\pm70\degree$ for the
  same mode sets as in Fig. 3.}
\label{fig:uxanti}
\end{figure}

Apart from the systematic shift in the meridional flow parameters globally
due to the correction of the position angle, the structure of the mean
flow parameters at each location on the disk remain very stable throughout
the course of the mission (Figure 7).
Certain features, such as the apparent variation
from east to west of the flow parameters, can scarcely be real, and must
result from systematic effects in the interpretation of the measurements.
Others, such as the reversal in sign of the meridional flow parameter
near the poles, suggestive of a high-latitude counter cell, may be real.
Both of these features, east-west asymmetries, and a meridional sign
reversal at high latitudes, appear broadly in the fast fits as well as the
13-parameter ring fits, though with slightly different structure and
amplitude. The fact that there is an east-west variation in the year-to-year
changes in the zonal flow as well as in the mean values indicates that
there is a sensitivity effect. The apparent enhancement in the zonal flow
at the east limb is certainly consistent with a spurious measured ``flow''
inwards from limb to disk center, such as has been posited by \citet{Zhao2012}
for the time-distance measurements and demonstrated to be an expected
effect \citep{BS12}. It could explain the observed meridional flow
reversal at the poles, with a comparable magnitude of order 10 m-s$^{-1}$.
It is
not however balanced by a comparable reduction in zonal flow at the west
limb. Whether the high-latitude meridional flow parameter reversal does in
fact reflect a more-or-less permanent feature of the meridional circulation
near the poles cannot yet be established without a better understanding of
the spatial systematics in the analysis of the measurements.

The year-to-year variations in annual averages of the zonal flow at different
latitudes and Stonyhurst longitudes show significant variations at the 2--5
m-s$^{-1}$-yr$^{-1}$ level, consistent with the amplitude of the global
torsional oscillation
signal which is azimuthally averaged and (at least for helioseismic
measurements) symmetrized about the equator \citep{howeetal2013}.
Note that both the zonal rotational acceleration and deceleration rates
were substantially larger in the last year than the first, with a marked
hemispheric asymmetry in both the amplitude and the zonal extent of the
decelerated belts, and a bifurcation of the equatorial accelerated zone
into two, centered slightly to the south of the equator.

\begin{figure}
  \scalebox{0.7}
  {\includegraphics{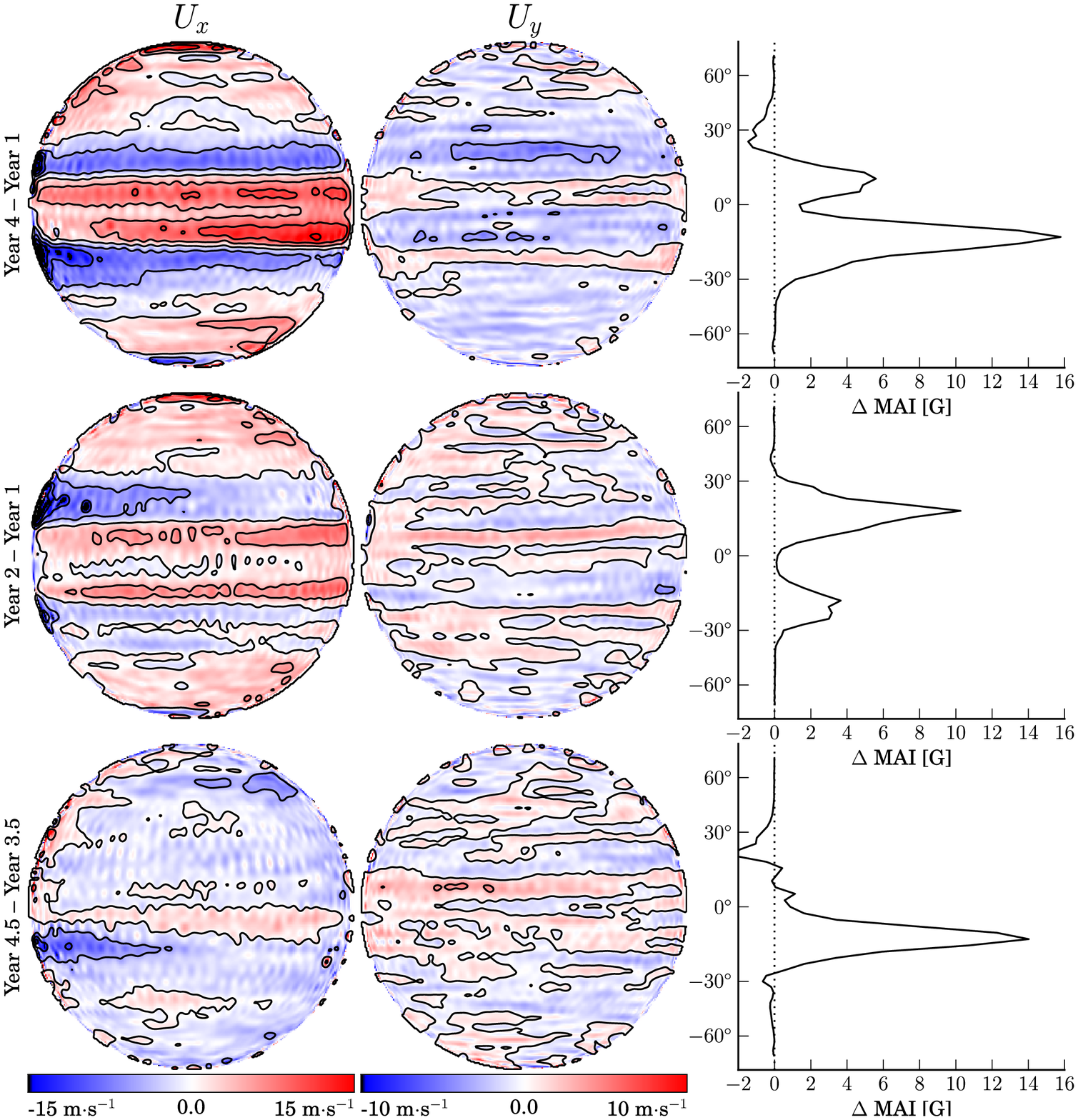}}
  \caption{
  Changes in the 1-year means of $U_x$ and $U_y$ between Year 1 and Year 4
  (top), Year 1 and Year 2 (middle), and Years 3.5 and 4.5 (bottom). The
  color scales range between $\pm15$ m-s${^-1}$ for $U_x$ and $\pm10$
  m-s$^{-1}$ for $U_y$
  with 5 m-s$^{-1}$ contours in both. For the year-to-year changes we also
  show the
  changes at each latitude of the mean values of the Magnetic Activity Index
  (MAI), defined as the average within each tracking cube of the absolute
  value
  of all HMI line-of-sight magnetic field values (from data series
  {\sl hmi.V\_45s}) in excess of 50 G.}
\label{fig:1yrchg}
\end{figure}

Variations in the mean meridional flow are more marginal. In the northern
hemisphere there has been a consistent strengthening in the poleward flow
of about 3 m-s$^{-1}$-yr$^{-1}$ in a narrow zone around latitude 7\degree.5
and a corresponding
weakening of about -1.5 m-s$^{-1}$-yr$^{-1}$ in a somewhat broader zone around
latitude
15\degree--20\degree. Trends in the southern hemisphere are more confused.
Strengthening
of the poleward flow centered around latitude -12\degree.5 in the first year
has migrated to around -30\degree\ in the last, while the zones of weakening
have correspondingly migrated toward the equator. (As discussed above, the
measured overall southward flow increase from the first year to the last is
only an artifact of the change in the assumed position angle used in the
tracking.)

Turning our attention to variations in the flow patterns that can be
localized in space and time, we note two striking features
({Figures 8 \& 9}). There are anomalies in the zonal flow at high latitudes
that manifest themselves along distinct bands in the Carrington coordinate
system of width about 10\degree.
They stretch from southwest to northeast in the northern hemisphere and
oppositely in the southern hemisphere, consistent with dragging of a feature
by the differential rotation shear.
They can be clearly seen in Figure 10, which combines the
standard synoptic maps with polar views for two selected rotations.
These features persist for several rotations, with
a gradual increase in their inclination angle, again consistent with shear
dragging. Both this winding and overall eastward propagation consistent
with surface differential rotation are more vividly illustrated in
Figure 11. Apart from these extended high-latitude features,
there are localized flow anomalies, equally visible in the meridional
and zonal directions, at lower latitudes that are clearly associated
with active regions, a well-established phenomenon; the anomalies in CR
2109 can be readily identified with AR 11190, 11193, 11195, 11199, 11203,
and 11204, while those in CR 2136 are associated with AR 11731 and
11734--11736.

The torsional oscillation in zonal flow anomalies at different latitudes
is fundamentally a function of two dimensions, time and latitude, as it
is based on azimuthal averages of the signal at each latitude over the
course of a rotation (or sometimes longer, as with the GONG, MDI, and
HMI helioseismic measurements based on 36- and 72-day averages ---
cf. Howe et al. 2013). Helioseismic measurements based on global
mode inversions also provide only the symmetric component of the
torsional oscillation signal over the northern and southern hemispheres.
Combining the present results which are spatially resolved in
two dimensions from the whole time range observed
allows us to produce a plot equivalent to standard plots of the torsional
oscillation, but with longitudinal resolution within each rotation and
the hemispheres separately resolved (Figure 12).

The localized small meridional flow anomalies, by contrast, exhibit
only occasional high-latitude structures, and they seldom last for
more than a single rotation. The anomalies are dominated by the
localized flows away from active regions. The equivalent plot to
standard torsional oscillation ones (Figure 12) exhibits alternating
bands of northward and southward flow anomalies at all latitudes, with a
period of one year. These are almost certainly due to an error in the
Carrington elements. Note especially that the amplitude of this annual
variation decreased (but did not completely disappear) at the time the
value of the HMI camera position angle with respect to the nominal
Carrington axis was corrected.

\begin{figure}
  \scalebox{0.65}
  {\includegraphics{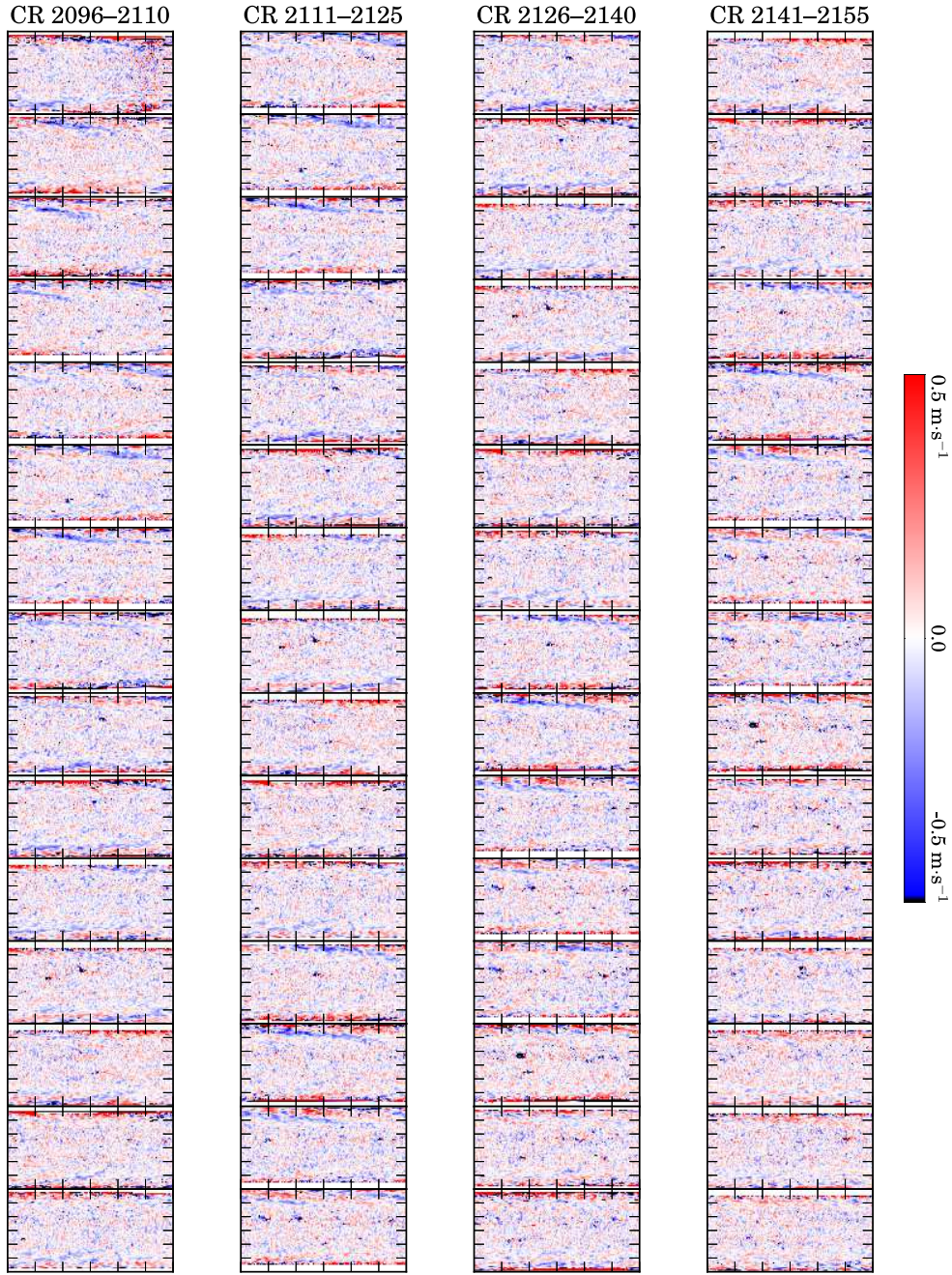}}
  \caption{
  Synoptic maps for each Carrington rotation of the $U_x$ flow anomalies,
  defined as the differences between the values at each heliographic location
  and the mean values at the corresponding Stonyhurst locations over
  Years 1--4,
  averaged over all observations during the rotation, for the same range of
  modes.  The color scale ranges between $\pm0.5$ m-s$^{-1}$, with red representing a
  positive (super-rotational) anomaly.
  }
\label{fig:synopuxanom}
\end{figure}

\begin{figure}
  \scalebox{0.65}
  {\includegraphics{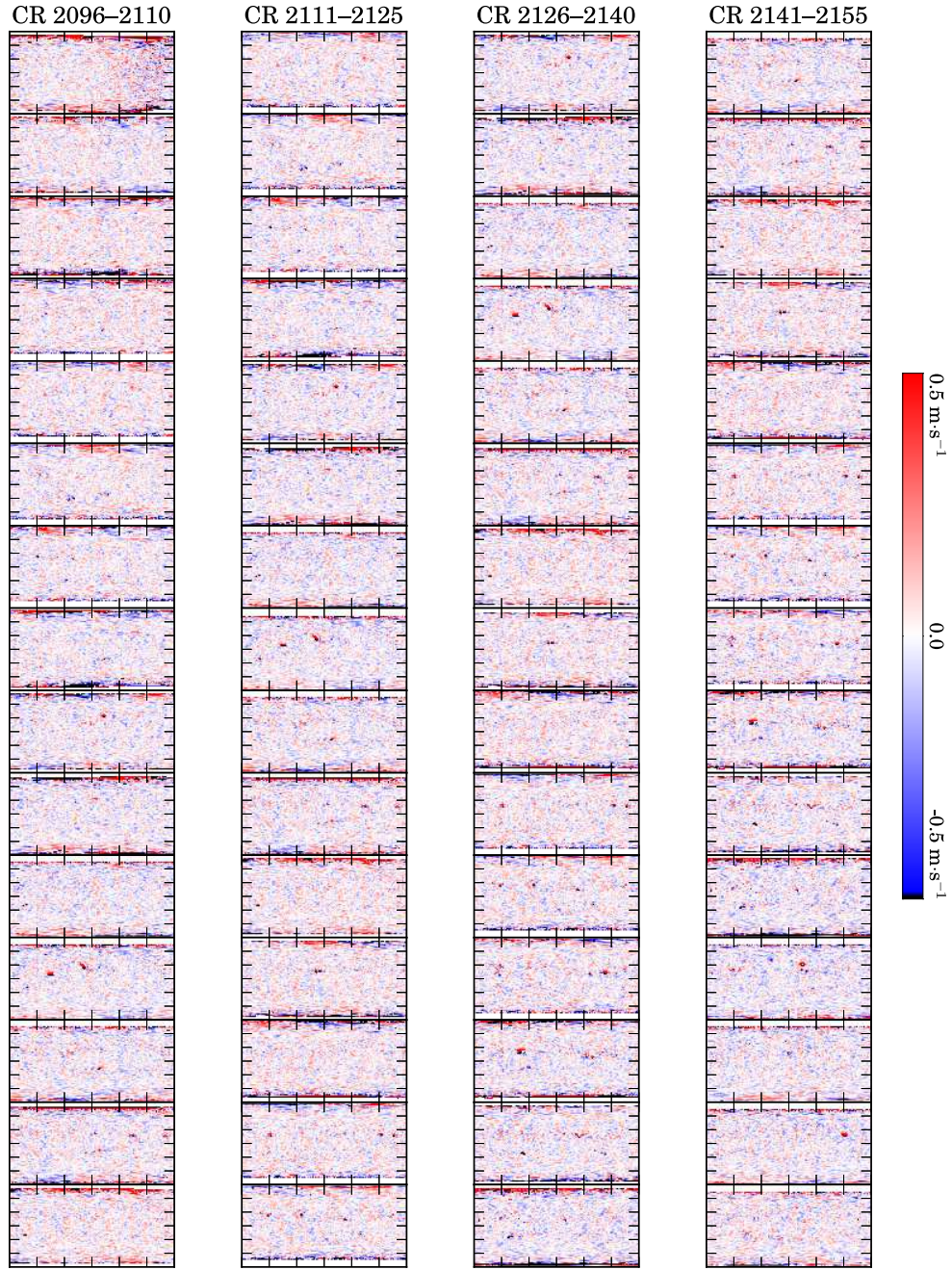}}
  \caption{
  Same as Figure 8, for the $U_y$ flow anomalies. The color scale ranges
  between $\pm0.5$ m-s$^{-1}$, with red representing a positive (northward) anomaly.}
\label{fig:synopuyanom}
\end{figure}

\begin{figure}
  \scalebox{0.70}{\includegraphics{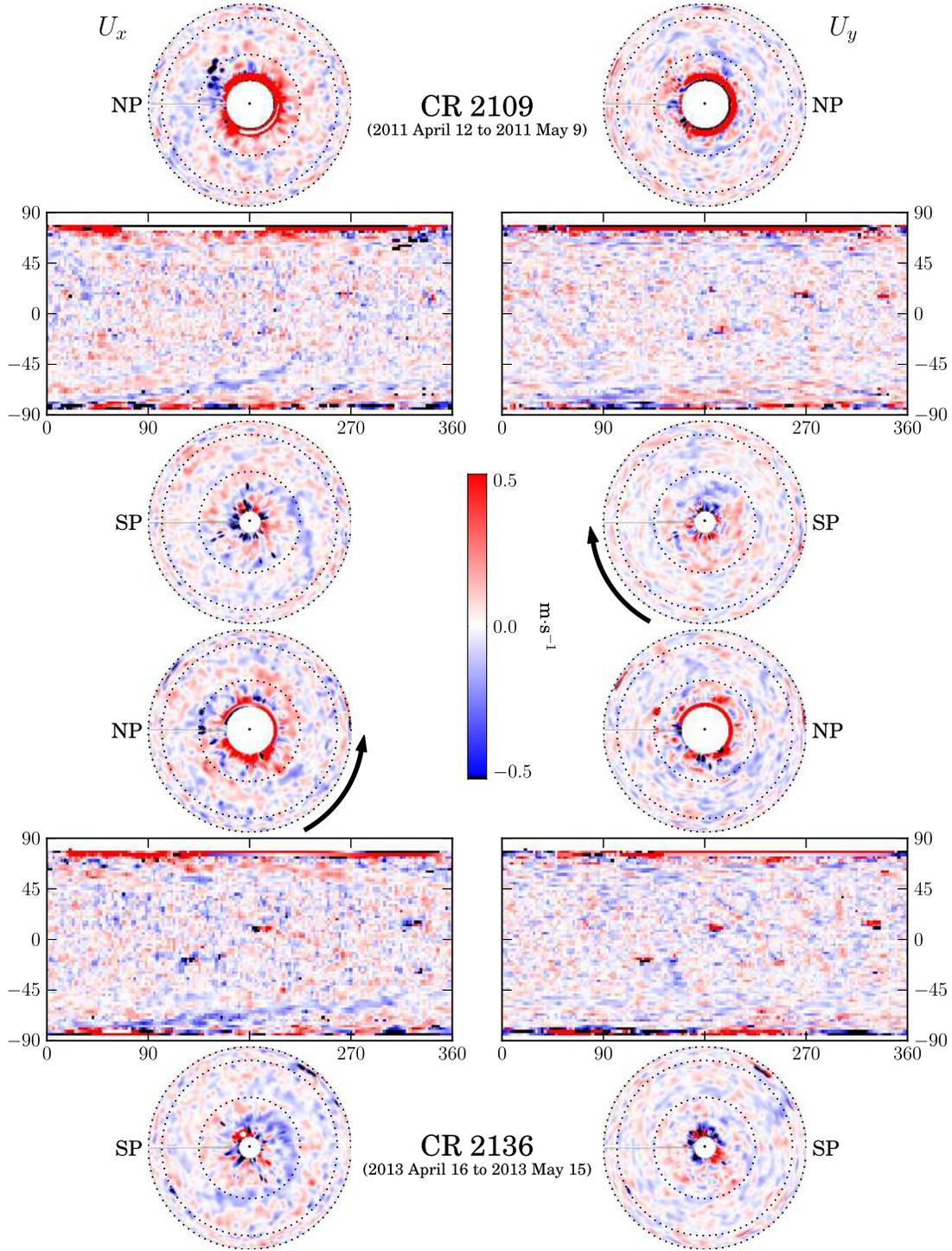}}
  \caption{
  Synoptic maps of the zonal flow anomalies $U_x$ (left) and $U_y$ (right)
  for two selected Carrington rotations two years apart, showing
  orthographic projections from above the north and south poles above
  and below the standard plate carr\'ee projections. The two
  rotations are at times when the south pole is tipped slightly toward the
  Earth, accounting for the more extended coverage at high southern latitudes.
  Carrington longitude 0\degree\ is at left in both polar plots, with
  longitude increasing in the direction of solar rotation as shown
  by the arrows.
  }
\label{fig:3rots3views}
\end{figure}

\begin{figure}
  \scalebox{0.75}
  {\includegraphics{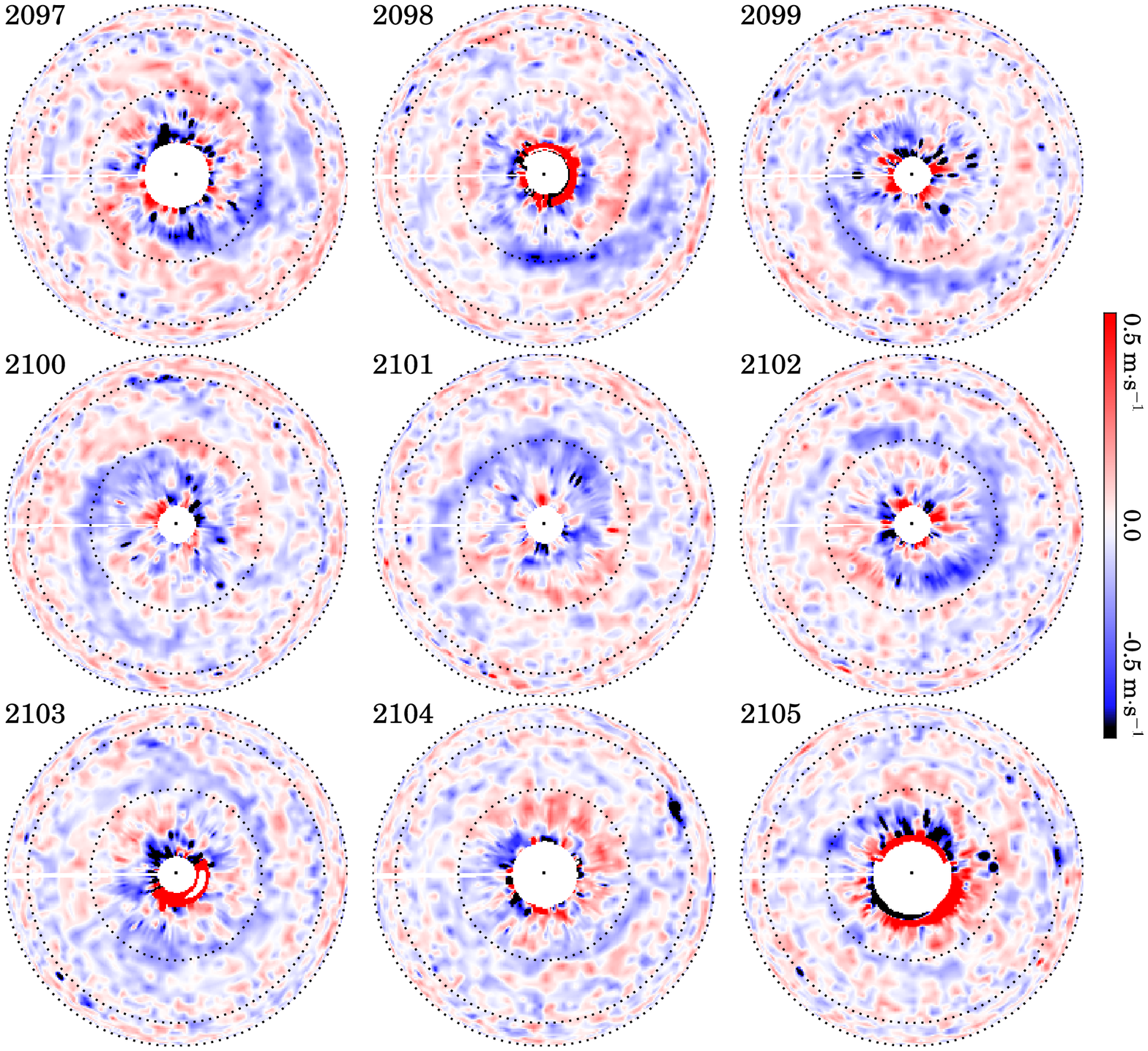}}
  \caption{
  Synoptic maps of the zonal flow anomalies in the northern hemisphere
  for nine successive Carrington rotations, plotted in an orthographic
  projection centered on the north pole. Carrington longitude 0\degree\
  is at left, with longitude increasing counter-clockwise, the direction of
  solar rotation. The mean sampling times of course increase clockwise,
  so the data from just below the 0\degree\ half-line are sampled one
  rotation later than the data just above. The color scale is the same
  as in Fig. 8. The dashed circles are latitude lines at intervals of
  30\degree\.\ The varying size of the unsampled region around the pole is
  due to the tipping of the pole toward and away from the Earth-orbiting
  observatory over the course of the year.}
\label{fig:polarux}
\end{figure}

Although the fitted mode sets from such small regions are too limited for
decent inversions of the depth structure of these features, we can get
rough estimates of their depths by simply dividing the mode sets into
groups with different values for the classical turning point $\nu / \ell$.
Figure 13
compares the values along cuts at selected latitudes from contiguous synoptic
maps of the anomalous zonal flows for three rotations, but constructed
for two different sets of modes, one with turning points at a
depth of about $4 \pm 1$ Mm and the other at about $6 \pm 1$ Mm. It is
evident that these anomalous zonal flow bands extend to depths
of at least 1\% of the radius. At the lower latitude there is no significant
difference between the anomalous flows at the two depths, both showing the
same longitudinal pattern. At the higher latitudes the anomalies clearly
have greater amplitude at the greater depth, although they are still
highly correlated, the peaks and valleys occurring at the same longitudes;
in all cases the probability of exceeding the correlations in random
distributions with the same statistics are far less than $10^{-3}$.
The greater amplitude of the
negative anomaly at the greater depth actually runs counter to the general
trend expected in the radial differential profile at high latitudes, but in
any case it is clear that its amplitude remains larger than any azimuthally
independent differences due to radial differential rotation.

\begin{figure}
  \scalebox{0.75}
  {\includegraphics{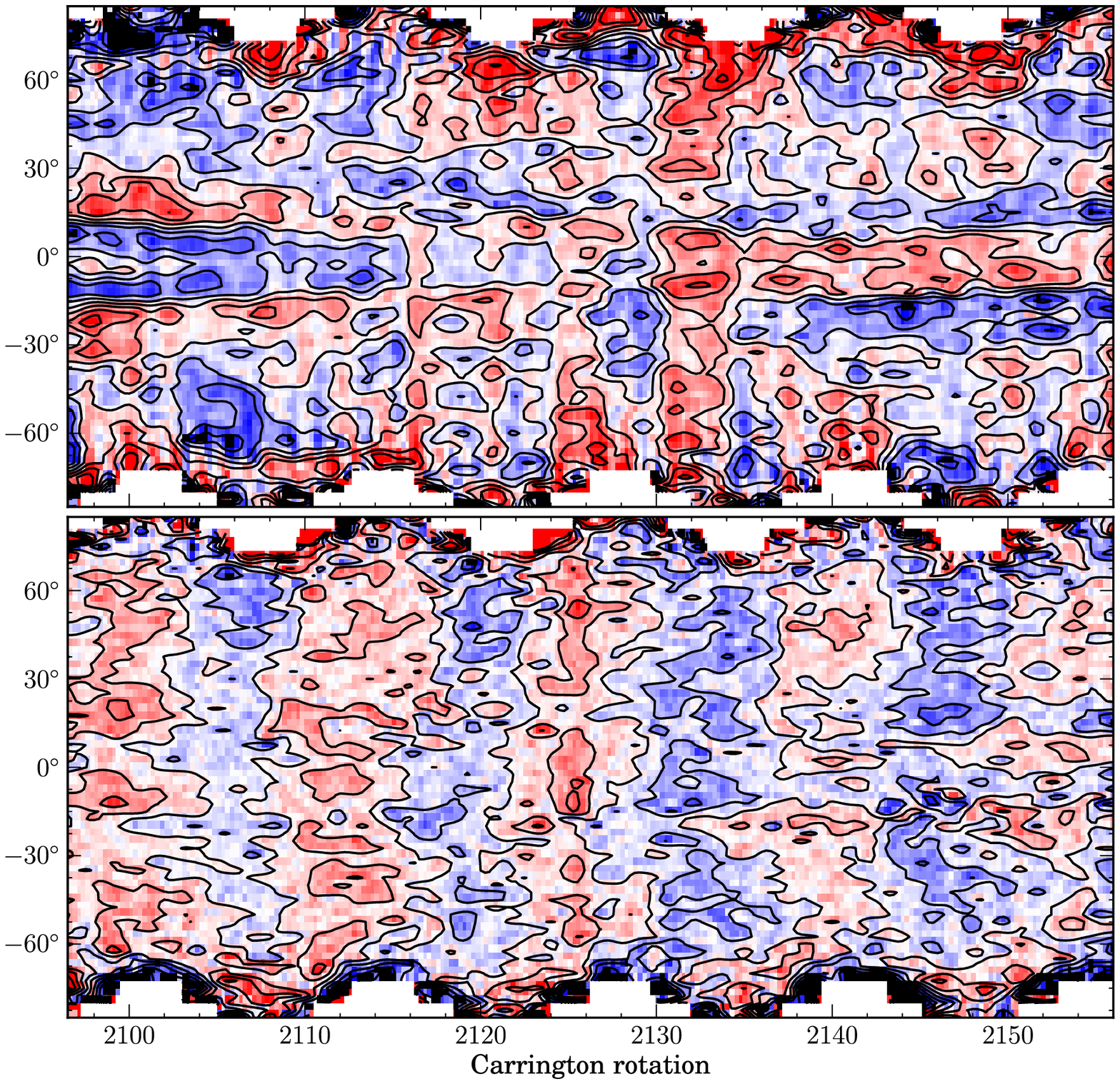}}
  \caption{Anomalies for the zonal (top) and meridional flow components
  relative to the 4-year means measured at the same disc positions, averaged
  over all longitudes and over a full Carrington rotation and sampled four times
  per rotation, plotted as functions of time on the horizontal axis and
  latitude on the vertical. The color scale range is $\pm10$ m-s$^{-1}$ for both,
  with contours at 2.5 m-s$^{-1}$ intervals. The time range covered extends from
  CR 2096.5 (1996 May 6) to 2056.0 (2014 October 14)}
\label{fig:torosc}
\end{figure}

\begin{figure}
  \scalebox{0.75}
  {\includegraphics{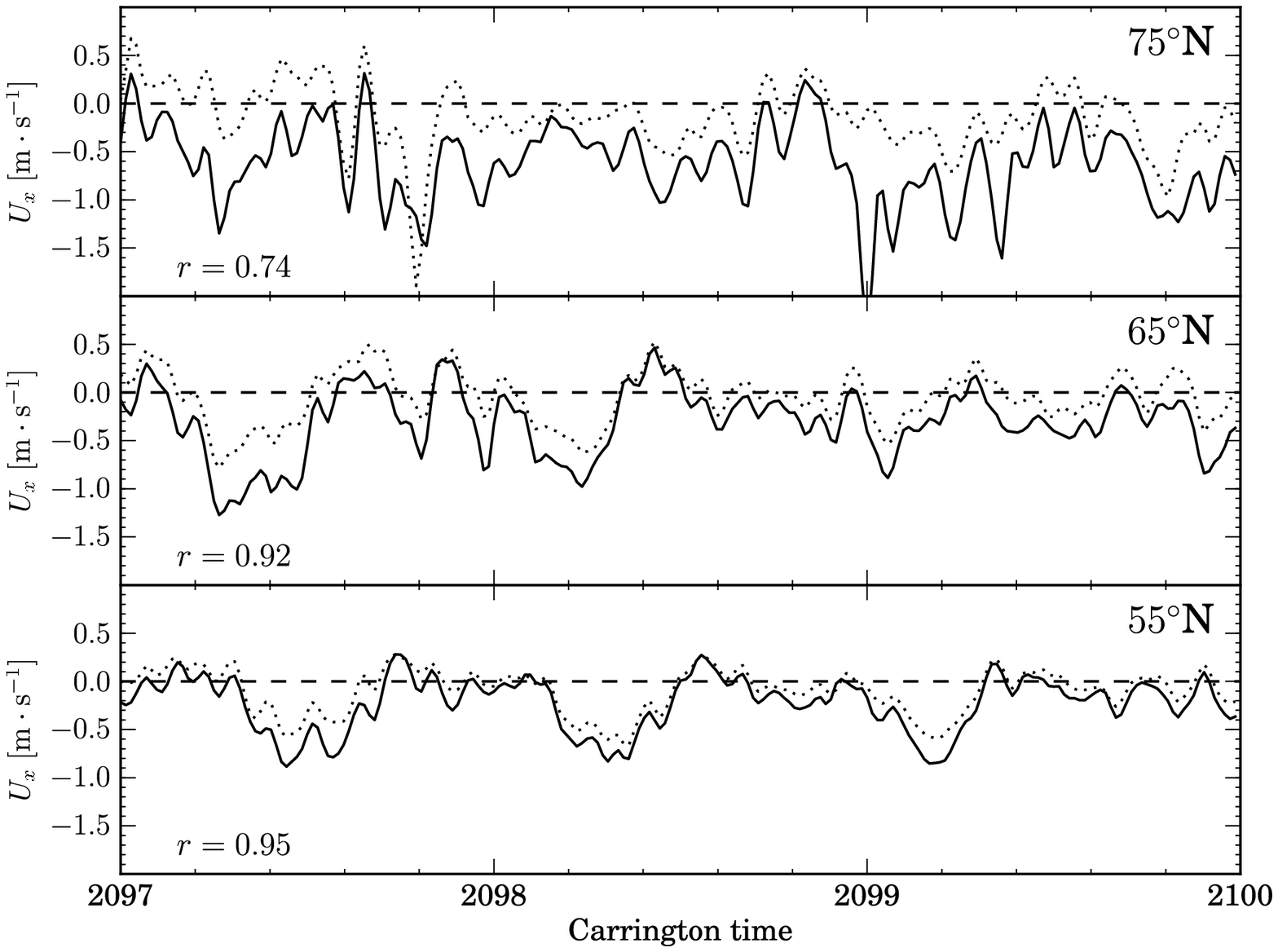}}
  \caption{Comparison of the zonal flow anomalies during CR 2097--2099 at
  selected latitudes for two different mode sets. The solid curve displays
  the local anomalies determined from modes restricted to those with classical
  turning points between 0.9900--0.9925 (depth 5.25--7 Mm), the dotted curve
  to those in the range 0.9925--0.9950 (depth 3.5--5.25 Mm). Each set of curves
  represents the mean zonal flow anomalies as a function of longitude within a
  latitude strip of width 10\degree\ centered on the labeled latitude. They are
  equivalent to plots of the values along horizontal cuts at different latitudes
  in a series of three consecutive synoptic plots of {Figure 8.} {\sl r} is the
  correlation coefficient between the flow anomalies at the different depths}
\label{fig:depthdep}
\end{figure}

\section{Conclusions}
Ring-diagram analysis of high-resolution Doppler data from HMI can resolve
features in the near-surface flow fields with limited depth resolution
at size scales down to about 50 Mm and extending almost to the poles.
There is some evidence for a regular counter-cell near the photosphere
at very high latitudes in the mean global meridional circulation, but
absolute determinations of mean flows at high latitudes are compromised
by systematic effects on both the measurements and analysis at large
center-to-limb angles that are not yet fully understood.

Despite the uncertainties in determining mean flows globally, it is evident
from several years' data that the systematic center-to-limb effects are stable.
By removing the measured means for observations at each disc position, and
carefully averaging over entire years so that all values of the heliographic
observer latitude are equally sampled, we can determine residuals in the
zonal and meridional velocities with a spatial resolution down to that of
the ring analysis itself, and a temporal resolution of less than or order
of a single Carrington rotation. We find that there are belts of localized
anomalies in the zonal flow at high latitudes persisting for five or more
rotations. These belts are elongated in longitude, at least qualitatively
consistent with the shear expected from differential rotation of a feature
initially aligned along a meridian.

It is interesting to speculate on the nature of the persistent structures
seen in the spatial distribution of the flow anomalies, particularly
those of the zonal flow at high latitudes. It is tempting to interpret
these flow anomalies as the eastward or westward motions at the surface of
extended ``banana cells'' convective structures, as suggested by
\citet{Hathaway13}, that have been ``wound up'' by differential rotation.
Certainly the drift of the spiral pattern in the direction opposite to
that of the rotation of the Carrington frame, especially evident in Figure
11, is to be expected from the slower rate of surface rotation at high
latitudes. On the other hand, there is little evidence of tightening of
the spirals from one rotation to the next. Quantitative investigation of
the motions of these patterns in comparison with the known depth structure
of the differential rotation profile might yield information on the depth
at which they are seated, and whether they are indeed surface manifestations
of deep convective cells. A point to note is that although nearby zonal flow
anomalies of opposite sense would imply divergence or convergence at lower
latitudes, near the poles they represent instead a zonal shear between the
inner and outer arms of the spiral. This might be related to the development
of the torsional oscillation at high latitudes, with its implications for
evolution of the solar cycle.

Although the meridional flow anomalies do not exhibit nearly as marked nor
persistent structures at high latitudes, there appears to be a general
tenedency at lower latitudes for them to organize themseleves in about eight
to twelve alternating bands of northward and southward motions. Whether this
pattern is statistically significant remains to be established; if it is,
it would suggest a large-scale sectoral structure of cyclonic and anti-cyclonic
flows that should be related to the zonal flow anomalies at the higher
latitudes. One notable feature in the high-latitude patterns of the
meridional flows is that rather then being organized into spiral structures,
the tend to exhibit simply a single-celled azimuthal asymmetry. This is what
would be expected if there were a cross-polar flow. If such flows were
slightly displaced from the exact pole, their azimuthal averages might
account for the apparent counter cell in the mean meridional flow that
we observe at high latitudes.

\acknowledgments
We are grateful to David Hathaway and to an anonymous reviewer for
helpful comments and suggestions.
This research was supported in part by NASA Contract NAS5-02139 and in
part by NASA Grant NNX14AH08G, both to Stanford University.

\bibliography{bibfile}
\end{document}